# Optimal Joint Multiple Resource Allocation Method for Cloud Computing Environments

Shin-ichi Kuribayashi[1]

[1]Department of Computer and Information Science, Seikei University, Japan

*Abstract*: Cloud computing is a model for enabling convenient, on-demand network access to a shared pool of configurable computing resources. To provide cloud computing services economically, it is important to optimize resource allocation under the assumption that the required resource can be taken from a shared resource pool. In addition, to be able to provide processing ability and storage capacity, it is necessary to allocate bandwidth to access them at the same time.

This paper proposes an optimal resource allocation method for cloud computing environments. First, this paper develops a resource allocation model of cloud computing environments, assuming both processing ability and bandwidth are allocated simultaneously to each service request and rented out on an hourly basis. The allocated resources are dedicated to each service request. Next, this paper proposes an optimal joint multiple resource allocation method, based on the above resource allocation model. It is demonstrated by simulation evaluation that the proposed method can reduce the request loss probability and as a result, reduce the total resource required, compared with the conventional allocation method. Then, this paper defines basic principles and a measure for achieving fair resource allocation among multiple users in a cloud computing environment, and proposes a fair joint multiple resource allocation method. It is demonstrated by simulation evaluations that the proposed method enables the fair resource allocation among multiple users without a large decline in resource efficiency.

## 1. Introduction

Cloud computing services are rapidly gaining in popularity. They allow the user to rent, only at the time when needed, only a desired amount of computing resources (processing ability and storage capacity) out of a huge mass of distributed computing resources without worrying about the locations or internal structures of these resources [1]-[4]. The popularity of cloud computing owes to the increase in the network speed, and to the fact that virtualization and grid computing technologies have become commercially available. It is anticipated that enterprises will accelerate their migration from building and owning their own systems to renting cloud computing services because cloud computing services are easy to use, and can reduce both business costs and environmental loads.

The National Institute of Standards and Technology (NIST) identified four essential characteristics of cloud computing: on-demand self-service, broad network access, resource pooling, rapid elasticity and measured service [4]. For example, Amazon Elastic Compute Cloud (EC2) [2] is an example of HaaS (Hardware as a Service), which is a form of cloud computing. This could be a virtual machine rental service (similar to a conventional 'server rental' service).

Cloud computing services enable the user to use the required computing resources for the required period, thereby enabling the user to build a flexible information system economically. In addition, there are cases where processing ability and storage capacity are provided in combination with a broadband path in the form of what is called an ICT platform service (or converged service).

As cloud computing services rapidly expand their customer base, it has become important to provide them economically. To do so, it is essential to optimize resource allocation under the assumption that the required amount of resource can be taken from a common resource pool and rented out to the user on an hourly basis. In addition, to be able to provide processing ability and storage capacity, it is necessary to reserve simultaneously a network bandwidth to access them. Therefore, it is necessary to allocate multiple types of resources (such as processing ability, bandwidth, and storage capacity) simultaneously in a coordinated manner instead of allocating each type of resource independently. The amount of resource required and the period in which it is used are not fixed. They can vary greatly from user to user and from service to service.

This paper proposes an optimal resource allocation method for cloud computing environments. For the preliminary evaluation, this paper assumes the use of two types of resources: processing ability and bandwidth. In general, services can be classified into two categories: a non-delay system (loss system) and a waiting system. A non-delay system allocates a spare resource immediately to the user upon the arrival of the request, and rejects the request if there is no spare capacity. A waiting system allocates a spare capacity to users in the sequence in which their requests have arrived, instead of allocating resources immediately upon the arrival of a request. This paper assumes a service that runs as non-delay. This paper also assumes static resource allocation, which is the most basic form of resource allocation, although dynamic allocation, which uses process migration and bandwidth consolidation, can increase the utilization of resources.

Section 2 develops a resource allocation model for cloud computing environments, assuming both processing ability and bandwidth are allocated simultaneously to each service request and rented out on an hourly basis. The allocated resources are dedicated to each service request. Section 3 presents an optimal joint multiple resource allocation method, based on the resource allocation model in Section 2. Section 4 defines basic principles and a measure for achieving fair resource allocation among multiple users in a cloud computing environment, and proposes a fair joint multiple resource allocation method. Section 5 explains the related work. Finally, Section 6 presents the conclusions. This paper is the extension of previous studies that were focused on an all-IP network [12],[13].

## 2. Resource allocation model for cloud computing environments

As mentioned in Section 1, the resource allocation in a cloud computing environment can be modeled as allocating the required amount of multiple types of resource simultaneously from a common resource pool for a certain period of time for each request. The allocated resources are dedicated (not shared) to each request. For the preliminary evaluation, this paper assumes two types of resource: processing ability and bandwidth.

The required amount of resource and the period of time in which it is used are not fixed. They can vary greatly from user to user and from service to service. For example, file transfer, video delivery, and videoconferencing services require a large amount of bandwidth but not so much processing ability. In contrast, an accounting service requires a large amount of processing ability but not so much bandwidth. It is assumed that the hardware resources for cloud computing services are not installed at a single center, but in multiple geographically distributed centers, as shown in Figure 1, in order to facilitate addition of resources, to implement load balancing and to ensure high reliability. Each center has servers (including virtual servers) that provide processing ability, and bandwidths that provide access to these servers. When a new request is generated, one center from among k centers is selected according to the resource allocation algorithm. The maximum size of processing ability and bandwidth at center j (j=1,2,...,k) is assumed to be $C_{maxj}$ and $N_{maxj}$ respectively. Figure 2 illustrates the concept of resource allocation that takes the resource usage period into consideration. When a service request arrives, the optimal center is selected from multiple centers, and both the processing ability and bandwidth available in the selected center are allocated for a certain period of time. The processing ability and bandwidth of only this single center are allocated. If no center has an adequate amount of spare resources (both processing ability and bandwidth), the request is rejected. The resources allocated are released after the usage period has elapsed.

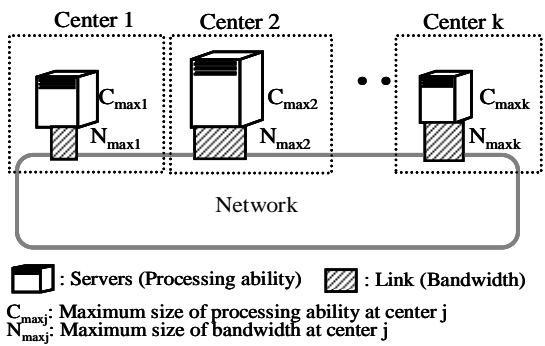

Fig. 1　System model for cloud computing services

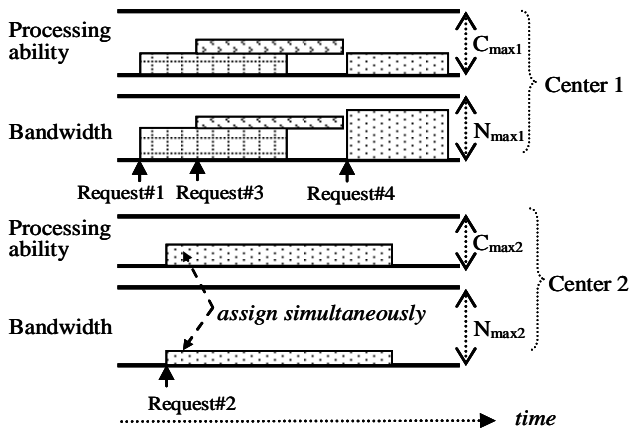

Fig. 2　Example of resource allocation for Fig.1 system model (k=2)

## 3. Optimal joint multiple resource allocation for cloud computing environments

### 3.1. Conditions

This section assumes non-delay system and static resource allocation, as discussed in Section 1. The objective of the allocation is to maximize the number of requests to which both processing ability and bandwidth are allocated well.

### 3.2. Impact of joint multiple resource allocation

As the size of required processing ability does not normally have a fixed relationship with that of required bandwidth, the optimal resource allocation can not be achieved if only a single resource type is considered in the selection of a center. Case 1 in Figure 3 illustrates an example of resource allocation in the case where only processing ability is considered. Case 1 tries to select the center according to Best-Fit approach, i.e. the center with the less available amount of processing ability is selected. It is clear that 'deadlock state', in which it is not possible to support a request that needs both types of resource even though spare resources of both types are available in the total sysytem, occurred in case 1 and an additional request ⑤ is rejected. This is because that case 1 does not consider bandwidth in the selection of a center and it is not an optimal allocation for bandwidth. In contrast, case 2 in Figure 3 illustrates an example of resource allocation in the case where both processing ability and bandwidth are considered in the selection of a center. The same request generation pattern as case 1 is assumed. It is clear that the deadlock state did not occur in case 2 and an additional request ⑤ is accepted.

Therefore, it is necessary to consider both processing ability and bandwidth in the selection of a center, and to study a new method for joint multiple resource allocation.

### 3.3. Optimal joint multiple resource allocation method

In general, it is difficult to consider multiple types of resource at the same time. It is proposed to identify one single resource type which has the largest impact on the resource allocation. We propose a new resource allocation method (**Method II**) which considers only 'identified resource' in the selection of a center. Method II adopts Best-Fit approach and aims to reserve as much as possible for future requests that may require a larger size of processing ability or bandwidth, and to reduce the possibility that the deadlock situation shown in Figure 3 will occur. As for comparison, Round Robin method (**Method I**) in which the center is selected in turn in the pre-defined order, is also considered here. This is one of the conventional methods, which does not consider the situation of both processing ability and bandwidth in the resource allocation.

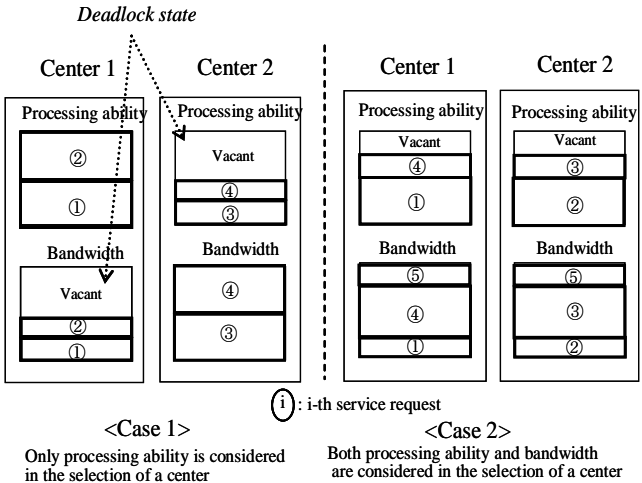

Fig. 3 Joint multiple resource allocation (k=2)

*3.3.1 Resource allocation by Method II*

The resource type that requires the largest proportionate size of resource, comparing the size of required resource with the maximum resource size for each resource type, is first selected as '**identified** resource'. Then the center with the least available amount of the identified resource from among k centers is selected. In this case, the units of processing ability and bandwidth are different, being measured in percentage of CPU power and b/s (bits per second) respectively. It is proposed to compare the size of the different resources as in the following example. Suppose that the maximum amount of bandwidth in a center is 100Mb/s. A request for 20% of CPU power and 30Mb/s requires 20% of processing ability and 30% of bandwidth respectively. As the proportion of required bandwidth is larger than that of required processing ability, bandwidth will be the identified resource in this case.

When a service request is generated, the following resource allocation algorithm is carried out:
i) Selection of identified resource
  **if** $X_C > X_N$ then processing ability is the identified resource.
  **else** bandwidth is the identified resource.
  where
   $X_C$={the size of required processing ability }/$X_{C0}$
   $X_{C0}$= Min {the maximum amount of processing ability in a center}
   $X_N$= {the size of required bandwidth}/$X_{N0}$
   $X_{N0}$= Min {the maximum amount of bandwidth in a center}

For example, if there are two centers and the maximum amount of processing ability of each center is 100 and 50 respectively, $X_{C0}$ will be 50.
ii) Selection of the center

The center which satisfies the following three conditions will be selected:
  - Min {the available size of the identified resource in the center}
  - Available processing ability in the center is equal to or larger than the required processing ability.
  - Available bandwidth in the center is equal to or larger than the required bandwidth.

If there are two or more centers which satisfy the above conditions, one center will be selected at random. Note that the request will be rejected if there are no centers that satisfy the above conditions.
iii) Allocation of resource
  Both required processing ability and bandwidth are allocated simultaneously in the selected center. The allocated resources are dedicated (not shared) to the request.
iv) Release of resource
  When the service time has expired, both allocated processing ability and bandwidth will be released simultaneously.

*3.3.2 Resource allocation by Method I*

When a service request is generated, the following resource allocation algorithm is carried out:
i) Selection of the center
  First, one center from among k centers is selected in a round-robin fashion. If there are not enough resources available for the selected center, the next center in the pre-defined order will be selected. For each request, the center which is next in the pre-defined order will be checked first, regardless of which center was selected by the previous request. Note that the request will be rejected if none of the centers have enough resources available.
ii) Allocation of resource
  Both required processing ability and bandwidth are allocated simultaneously in the selected center. The allocated resources are dedicated (not shared) to the request.
iii) Release of resource
  When the service time has expired, both allocated processing ability and bandwidth will be released simultaneously.

**3.4. Simulation evaluations**

*3.4.1 Simulation conditions*
1) The evaluation is performed by a computer simulation using the C language. The simulation model is based on the case where k is 2 in Figure 1. That is, there are two centers (center 1 and center 2). The maximum size of processing ability of center 1 and center 2 is assumed to be $C_{max1}$ and $C_{max2}$ respectively. The maximum size of bandwidth of center 1 and center 2 is assumed to be $N_{max1}$ and $N_{max2}$ respectively. When a service request is generated, center 1 or center 2 is selected according to the resource allocation algorithms proposed in Section 3.3.
2) The size of required processing ability and that of bandwidth follow a Gaussian distribution, and average values are given by C and N respectively.

3) The generation interval of requests follows an exponential distribution (average interval of request arrival is given by q). The service time H, which is the total time from a generation of the request to a completion of the service, is assumed to be constant. Each request occupies the allocated processing ability and bandwidth until its service time H passes.

4) It is supposed here that the following m requests will be generated repeatedly. The value of C and N of the first generated request are $a_1$ and $b_1$ respectively. The value of C and N of the second generated request are $a_2$ and $b_2$ respectively. The value of C and N of the m-th generated request are $a_m$ and $b_m$ respectively. We denote this request generation pattern like $\{C=a_1,N=b_1;C=a_2,N=b_2;…;C=a_m,N=b_m\}$ in this paper.

*3.4.2 Simulation results and analyses*

Figures 4, 5 and 6 illustrate the result of simulations which compare Method II with Method I. Figure 4(1) compares the request loss probability, which is the probability that either processing ability or bandwidth is not available, in the case where the value of C is the same as that of N. Figure 4(2) compares the request loss probability in the case where the sizes of processing ability and bandwidth (C and N) rise and fall in anti-phase, i.e., a large processing ability is followed by a large bandwidth. Figure 5 evaluates the impact of ratio of maximum resource size of each center on the request loss probability, assuming the total size of processing ability ($C_{max1}+C_{max2}$) and that of bandwidth ($N_{max1}+N_{max2}$) are constant. Figure 6 evaluates the impact of the number of centers on the request loss probability with the same simulation conditions as those of Figure 4(2).

The following points are clear from these Figures:

(1) Method II can reduce the request loss probability and as a result, reduce the total amount of resource compared with Method I, in the case where the sizes of processing ability and bandwidth rise and fall in anti-phase. This is also true even if the number of centers increases except for the case where the number of centers odd.

This is because the deadlock state as in Figure 3 occurs readily in Method I when the number of centers is odd and the sizes of processing ability and bandwidth rise and fall in anti-phase. It is noted that the request loss probability will be small when the number of centers increases because the number of centers in the state that can be processed increases relatively for the same amount of service demands.

(2) To allocate resources to one specific center as much as possible is better rather than to distribute resource to multiple centers if the total amount of resource is constant. For example, the request loss probability when $C_{max1}$ and $N_{max1}$ are 40 (the total resource is allocated only to center 1) is small compared with the request loss probability when $C_{max1}$ and $N_{max1}$ are 20 in Figure 5.

## 4. Fair joint multiple resource allocation

There is a possibility that unfair allocation of resources, in which most of resources are occupied by the specific user, occurs with Method II. Therefore, this section proposes to enhance Method II to enable the fair resource allocation among users.

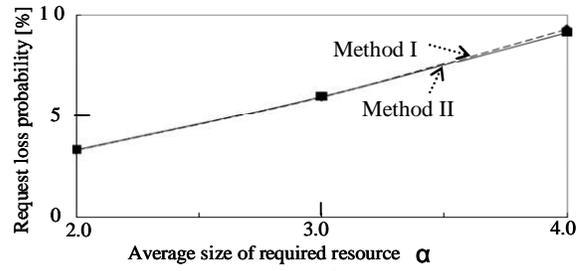

(1) Request generation pattern 1
$\{C=\alpha, N=\alpha\}$

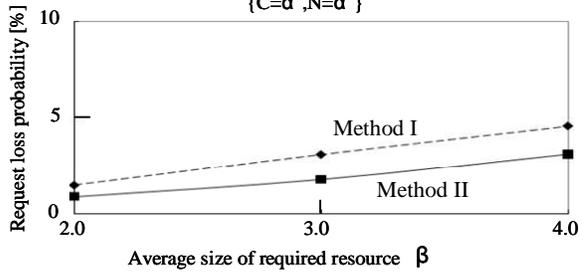

(2) Request generation pattern 2
$\{C=\beta, N=1; C=1, N=\beta\}$
$C_{max1}=C_{max2}=20$, $N_{max1}=N_{max2}=20$, $H=6$

Fig. 4 Comparative evaluation of Method I and Method II

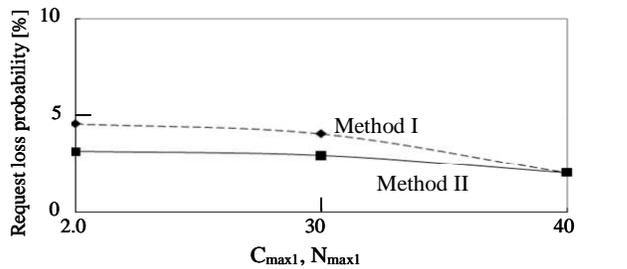

$C_{max1}+C_{max2}=40$, $N_{max1}+N_{max2}=40$, $H=6$ $\{C=4,N=1; C=1,N=4\}$

Fig. 5 Impact of the size of $C_{max}$, $N_{max}$

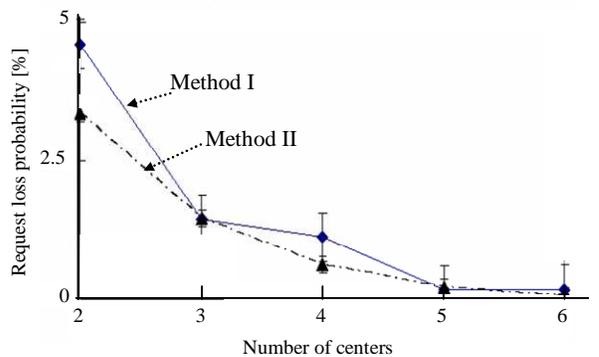

$C_{max1}=C_{max2}=20$, $N_{max1}=N_{max2}=20$, $H=6$; $\{C=4,N=1;C=1,N=4\}$

Fig. 6   Impact of number of centers

### 4.1. Basic principles for achieving fairness in joint multiple resource allocation

The following four basic principles are proposed to achieve 'fairness' among users in the joint multiple resource allocation:

<Principle 1> Since this paper assumes the non-delay resource allocation, it aims to achieve fairness without queuing in principle. However, it allows a delayed allocation of resources to serve some requests if it is necessary to achieve fairness.

<Principle 2> Fairness should be pursued while taking multiple types of resource into consideration. There are many papers that discuss algorithms for achieving fairness for cases where a joint resource allocation is not considered [5]-[9]. These algorithms cannot be applied to cases where multiple types of resource are allocated simultaneously, because an allocation that is fair for one specific resource type may be unfair when considering other resource types. This is caused by the fact that the size of the requested processing ability and that of the requested bandwidth are not generally uniform and both processing ability and bandwidth should be allocated simultaneously.

<Principle 3> Equalizing the total amount of both processing ability and bandwidth for all users would not achieve fairness. A possible solution to satisfy principle 2 may provide an equal amount of both processing ability and bandwidth to each user. However, this solution has a problem, as illustrated in Figure 7. It is supposed in Figure 7 that there are two users and half the amount of processing ability and bandwidth combined is a maximum allowable size to each user. If resources are allocated to one user like pattern 2, it will take up almost all bandwidth although the total amount of allocated resources to the user does not exceed the maximum allowable size. In this case, it is impossible to meet requests from the other user that may have pattern 1 or pattern 2 (both patterns require a large amount of bandwidth).

<Principle 4> It is not unfair when there are no request rejections occurred for all users, even if there is an imbalance in the amount of allocated resources to each user.

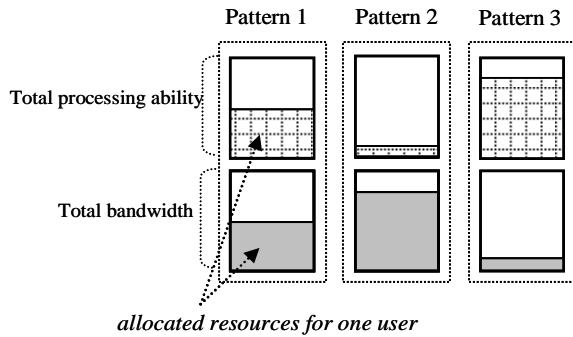

Fig.7 Possible resource allocation patters for each user

### 4.2. Proposed joint multiple resource allocation for achieving fair allocation

*4.2.1 Definition of 'fairness'*

This paper defines that the fairness can be achieved by allocating resources to each user in every time block, in proportion to the expected amount of resources requested by each user. The total amount of resources allocated to each resource type in a time block of length L is calculated, and then the resource type for which the total amount of requested resource is largest portion of the maximum resource of that type is selected as '**key** resource'. The fairness will be judged with the total amount of allocated key resource in a time block. This idea is very similar to the idea of Method II in Section 3. Key resource could change over time and could be different for each user. Note that L is assumed to be longer than resource holding time H.

*4.2.2 Measure of fair allocation*
*(a) Conditions*

 1) The number of users: G (user 1～user G)
 2) The ratio of expected amount of key resources requested by user g (g=1～G) to the least expected amount of key resources requested by a users is $1/r_g$.
 3) The user which has the largest value of $V_j(g)$ is called as 'user $g_1$' in j-th time block, where $V_j(g)$ is given by {Total amount of key resource allocated to user g in i-th time block}*$r_g$. $r_g$ is multiplied to normalize the amount of allocated resource, as in Figure 8. The resource allocation will be considered as fair when the normalized amount of allocated key resource is the same among users.
 4) The difference between total amount of key resources allocated to user $g_1$ and those allocated to user g in j-th time block is set to $N_j(g)$ and is called as 'Imbalance on allocated resources of user g (g=2～G)', as shown in Figure 9. $N_j(g)$ is given by $V_j(g_1)-V_j(g)$ and $N_j(g_1)$ be set to 0.
 5) If no request of user g is rejected in j-th time block, it is considered that the resource allocation is not unfair for user g, and $N_j(g)$ is set to 0, even if there are some differences in the amount of allocated key resource.

*(b) Measure of fair allocation*

It is proposed to check the value F given by Equation (1) and judge that the smaller the value of F is, the fairer the resource allocation is:

$$F = \sum_{j=1}^{s} \left[ \left\{ \sum_{g=1}^{G} N_j(g) \right\} / s \right] \quad (1)$$

where s is the total number of time blocks. Figure 10 illustrates the meaning of formula F.

If the value of F is the same for multiple users, it is proposed to judge that the smaller the change of the size of $N_j(g)$ is, the fairer the resource allocation is. The change of the size of $N_j(g)$, $F_1$, can be estimated with Equation (2):

$$F_1 = \sum_{j=1}^{s} \left[ \sum_{g=1}^{G} \{N_j(g) - N_{ave}(g)\}^2 \right] / s \quad (2)$$

where $N_{ave}(g)$ is the average of $N_j(g)$ in all time blocks.

*4.2.3 Fair joint multiple resource allocation method*

This section proposes a new joint multiple resource allocation method which is the extension of Method II and can achieve fair resource allocation. As this paper assumes the non-delay resource allocation, it is difficult to take any action in advance, avoiding any imbalance on the total amount of key resources allocated to different users. Therefore, it is proposed that the imbalance in j-th time block will be filling up in (j+1)-th time block, by applying the delayed resource allocation only to the user that suffered imbalance in j-th time block, on condition that the service shall be completed within a maximum permissible service completion time T. If there are

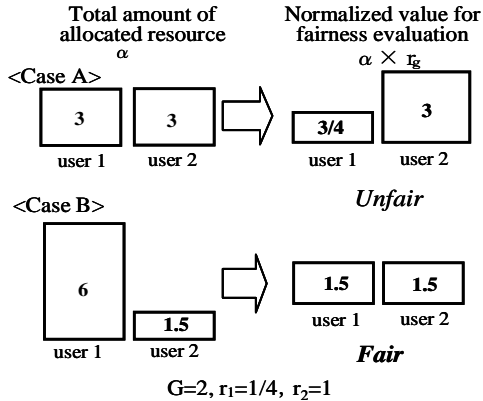

Fig. 8  Normalization of required resource requirement for fairness evaluation

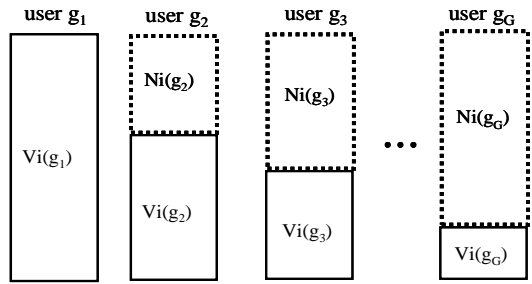

Fig. 9  Calculation of $N_j(g)$ in j-th time block

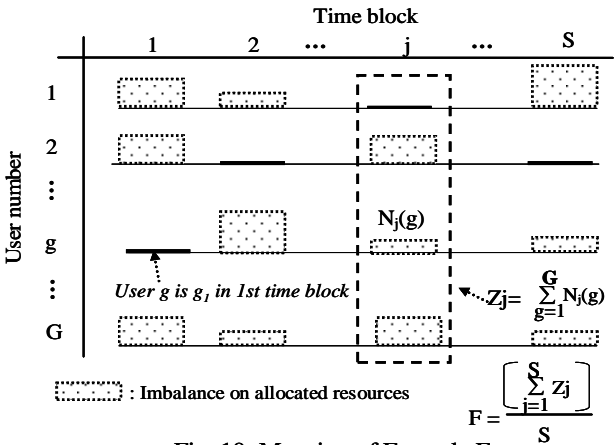

Fig. 10  Meaning of Formula F

Details of algorithm to fill up the imbalance in the next time block are as follows:

1) If requests for user g to which the amount of allocated key resource is small in j-th time block were rejected for a lack of resource in (j+1)-th time block, resources up to $N_j(g) / r_g$ will be filled up in (j+1)-th time block, regardless of whether requests for the user to which the amount of allocated key resource is large in j-th time block are rejected or not.

2) If no requests for user g to which the amount of allocated key resource is small are rejected for a lack of resource in j-th time block, it is considered that enough resource has been allocated to user g and no action is made to fill up the imbalance in (j+1)-th time block ($N_j(g)$ is set to 0), even if requests for the user to which the amount of allocated key resource is large in j-th time block are rejected.

3) When the delayed resource allocation is applied to fill up

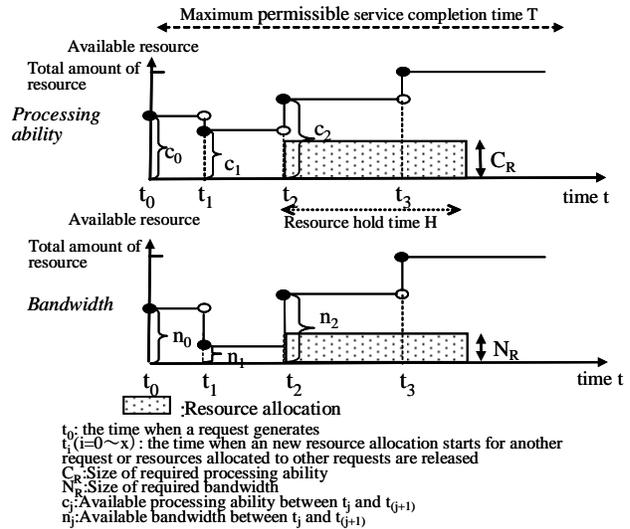

Fig. 11  Available resource management diagram (managed per each center)

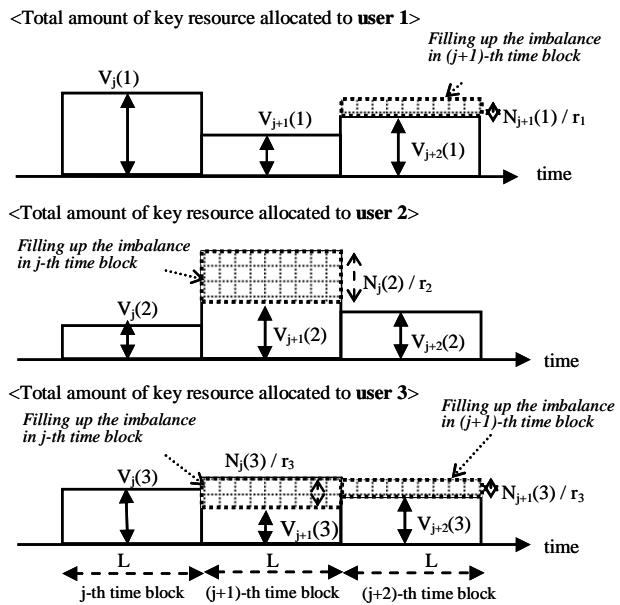

Fig. 12  Example of filling up the imbalance in the previous time block (G=3)

not enough resources available when a user request occurs, the resource allocation will be delayed until the required size of resources are available, instead of rejecting the request. The delayed resource allocation will be achieved until the total amount of key resource assigned for delayed requests exceeds $N_j(g) /r_g$ in (j+1)-th time block. It is impossible to decide which resource type is key resource before the end of (j+1)-th time block. Therefore, both $N_{j-c}(g)$, imbalance of processing ability, and $N_{j-n}(g)$, imbalance of bandwidth, are calculated from the beginning of each time block and the request will be rejected when $N_{j-c}(g) / r_g$ or $N_{j-n}(g) / r_g$ exceeds $N_j(g) / r_g$.

the imbalance, it is necessary to determine the time when to start the service. For this purpose, it is proposed to manage the available resource management diagram as illustrated in Figure 11. This diagram is managed per user.

Figure 12 illustrates an example of fill up the imbalance when G is three. We call the method which follows the above algorithm in unfair case but follows Method II in the fair case, as '**Method III**' in this paper.

### 4.3. Simulation evaluations

*4.3.1 Conditions*
All except for G=2 are same as the conditions in section 4.3.1.

*4.3.2 Simulation results and analyses*
Figures 13 and 14 show simulation results. The request generating pattern of user 1 and user 2 in those figures is indicated as {C=x, N=x} and {C=z*x, N=z*x}, respectively. {C=x, N=x} means that a request with C=x and N=x will be occurred repeatedly. z is the ratio of the size of user 2 request as opposed to the size of user 1 request. The average resource utilization in Figure 13 is the average usage of both processing ability and bandwidth. The following points are clear from these figures:
1) Method III can decrease the value of F greatly (that is, Method III enables fair resource allocation), compared with Method II which does not consider fair allocation. This is also true even if the number of centers increases.

In Figure 13, the value of F for Method II becomes large first as z becomes large, because the imbalance of the amount of allocated resource becomes large. However, where z exceeded 3.0, the value of F for Method II becomes small as z becomes large, because the size of required resource for user 2 requests is too large and it will be difficult to get enough resources available. On the other hand, the value of F for Method III increases a few as z becomes large.
2) Method III enables fair resource allocation in proportion to the expected amount of resources requested by each user, as in Figure 14.
3) In general, the resource efficiency will fail if we pursue fairness. However, Method III can attain the almost same resource efficiency as Method II, as in Figure 13.

## 5. Related Work

The proposed resource allocation model in Sections 2, 3 and 4 is similar to the online bin packing problem [5]. The bin packing problem is a problem of determining how items can be put in multiple bins most efficiently. In an online problem, the information about items is not known in advance. The proposed resource allocation model differs from the conventional bin packing problem in the following points:
  1) It deals with "composite bins", in which two bins (a processing ability bin and a bandwidth bin) are consolidated.
  2) Each item has two sizes, and they are simultaneously packed in composite bins.
  3) Packed items are removed after a certain period has elapsed. This period differs from item to item.
 The conventional bin packing problem is an NP-complete

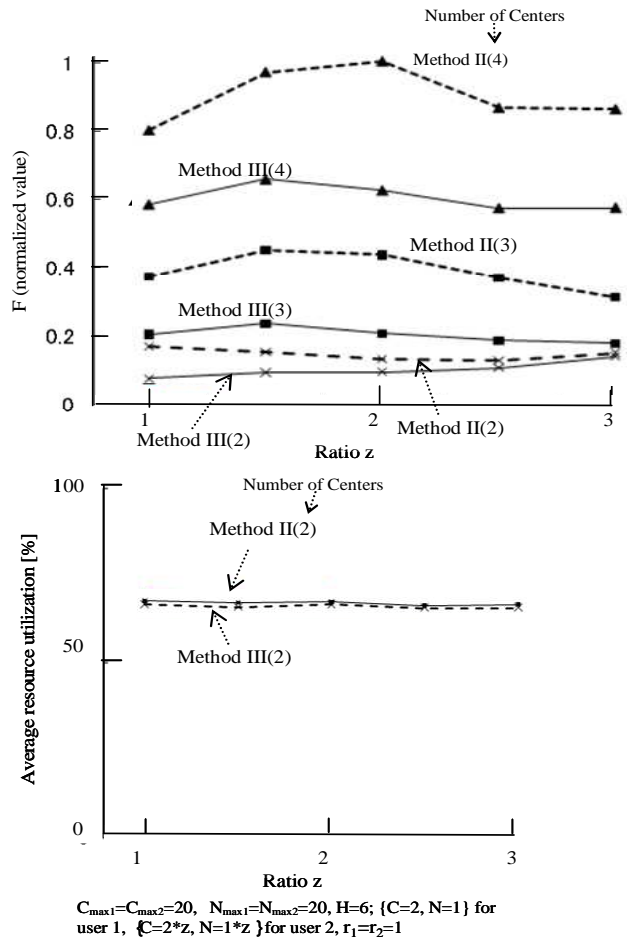

Fig. 13 Evaluation of fairness (value F) and resource efficiency

$C_{max1}=C_{max2}=20$, $N_{max1}=N_{max2}=20$, H=6; {C=2, N=1} for user 1, {C=2*z, N=1*z} for user 2, $r_1=r_2=1$

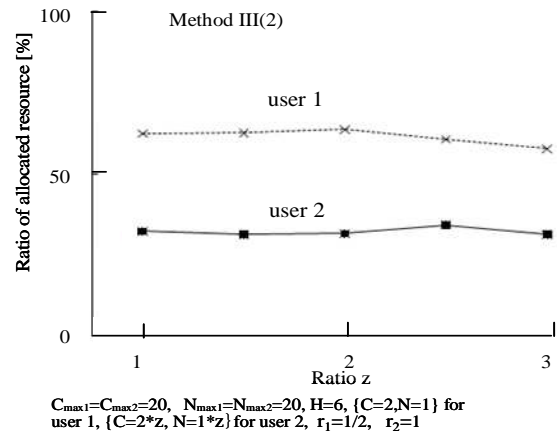

$C_{max1}=C_{max2}=20$, $N_{max1}=N_{max2}=20$, H=6, {C=2,N=1} for user 1, {C=2*z, N=1*z} for user 2, $r_1=1/2$, $r_2=1$

Fig. 14 Evaluation of ratio of allocated resource

problem, for which it is difficult to derive a general solution. Among the conventionally proposed algorithms for the online pin packing problem [5] Best-Fit (select bins with a smaller spare space) is considered to be more effective than Next-Fit, Worst-Fit, or First-Fit. We proposed an algorithm that extends the consolidated bin packing problem assuming Best-Fit.

Moreover, server selection methods that take both the server (processing ability) and the network (bandwidth) into consideration have been evaluated, as in Reference [11]. However, these methods assumed that each client requests a job processing and both servers and network are shared with multiple jobs from many clients in which servers and networks are modeled as queuing systems, and do not cover the model in which both processing ability and bandwidth, dedicated to each request, are rented out simultaneously on a hourly basis.

## 6. Conclusions

This paper has evaluated the resource allocation method for cloud computing environments.

This paper first has developed a resource allocation model for cloud computing environments and has proposed the optimal joint multiple resource allocation method, Method II, assuming that both processing ability and bandwidth are allocated simultaneously for each request and rented out on an hourly basis. The allocated resources are dedicated to each service request. It has been demonstrated by simulation evaluation that Method II could reduce the request loss probability and as a result, reduce the total amount of resource, compared with the conventional allocation method.

Then, this paper has proposed basic principles and a measure for achieving fair resource allocation among multiple users in a cloud computing environment, and has proposed a fair joint multiple resource allocation method, Method III, which tries to allocate resources in proportion to the expected amount of resources requested by each user. It has been demonstrated by simulation evaluations that Method III enables the fair resource allocation among multiple users without a large decline in resource efficiency, compared with the conventional method which does not consider the fair allocation.

## Acknowledgment

We would like to thank Mr. Shigehiro Tsumura and Mr. Kenichi Hatakeyama for their help with the simulation.